\documentstyle[preprint,eqsecnum,aps]{revtex}
\begin{document}
\draft
\tighten
\preprint{\vbox{\baselineskip=12pt
\rightline{SU-GP-98/5-1}
\rightline{hep-th/9805121}}}
\title{Localized Branes and Black Holes}
\author{Sumati Surya}
\address{I.U.C.A.A., Post Bag 4, Ganeshkhind, Pune 411007, India}
\author{Donald Marolf}
\address{Physics Department, Syracuse University, Syracuse, New York 13244}
\date{May, 1998}

\maketitle

\begin{abstract}
We address the delocalization of low dimensional D-branes and NS-branes
when they are a part of a higher dimensional BPS black brane, and the
homogeneity of the resulting horizon.  We show that the effective
delocalization of such branes is a classical effect that occurs when
localized branes are brought together.  Thus, the fact that the few known
solutions with inhomogeneous horizons are highly singular need not indicate
a singularity of generic D- and NS-brane states.  Rather, these singular
solutions are likely to be unphysical as they cannot be constructed from
localized branes which are brought together from a finite separation.
\end{abstract}

\vfil
\eject

\newcommand{\pd}{\partial}
\newcommand{\bpd}{{\bar \pd}}
\newcommand{\cF}{{\cal F}}
\newcommand{\R}{{\rm I\!\rm R}}
\baselineskip = 16pt

\section{Introduction}

The construction of BPS black brane and other solutions
\cite{HS,Duff,CT2,CT,AT,BDL,AT2,G} from D-branes and NS-branes has been
invaluable in studying black hole entropy
\cite{SV,CM,HS2,BMPV,MS,JKM,BLMPSV,KT,JM,DMW,DM1,DM2}, dualities
\cite{G,GGPT,AS}, and other phenomena within string/M-theory
\cite{BKOP,RPS}.  Such black $p$ branes are typically constructed from
several types of D- or NS-branes, including branes of dimension less than
$p$.  Thus, from the perspective of these lower dimensional branes, the
black $p$ brane has transverse internal directions.  It is an interesting
fact that all known smooth solutions have horizons which are
translationally invariant in the internal transverse directions
\cite{ght,HM1,HM2,HM3}.  Thus, all features of the black brane, including
the distribution of D- and NS- branes as well as any waves that may be
present (at least outside the horizon \cite{HM3}) will share this
invariance.  This is in contrast to the fact that, as long as the branes
remain separated (and properly oriented), one can construct static
solutions describing such D- and NS- branes distributed in an arbitrary
manner with respect to the internal directions.  Indeed, any solution in
which only a finite number of lower dimensional branes are present will not
be translationally invariant in any internal transverse direction.  Much
the same feature arises in the construction of intersecting brane solutions
\cite{BC,BMM,MM,BLL} at various angles, as those branes also appear in a
delocalized form.

It is important to note that related solutions {\it can} be constructed in
which this internal transverse symmetry is broken either by the distribution
of branes themselves or by various `waves' that are associated with the branes.
However, 
such solutions are highly singular \cite{HM2,HM3}.  Since a black brane
is thought to correspond to an ensemble of D-brane states, the assumption that
a generic bound state of D-branes would not have this internal
transverse symmetry leads naturally to the conclusion 
that such singular states will dominate the physics near the horizon
of a black brane \cite{KMR,RM,DM}.  It is therefore important to understand if
this assumption is correct.

We address this issue below by analyzing in detail classical BPS solutions
corresponding to collections of branes that are localized, but separated
from one another.  In particular, we consider the construction of BPS black
fivebranes in ten dimensional string theory from onebranes and fivebranes,
as is familiar from the study of black hole entropy for five dimensional
(three charge) black holes \cite{HS2}..  The distributions of onebranes and
waves are described by a collection of moments around an internal
transverse torus.  We study what happens when the separation between the
onebranes and fivebranes goes to zero while the intrinsic moments that
describe the distributions are held fixed.  One special case of this
process is when a fixed number of localized onebranes is brought into
contact with the fivebranes.  We find that the result is {\it not} a
singular solution.  Instead, as the infinite throat of the incipient black
hole forms and the onebranes move farther and farther inside it, the
various moments that describe the inhomogeneity of the solution are
`screened' from a distant observer.  As measured from the external region,
the multipole moments of the corresponding fields go to zero and the
solution goes over to the well known homogeneous one.  Thus, even if the
solution is constructed by moving a single localized onebrane onto a
fivebrane, the result is a smooth black hole solution with an exact
translational symmetry in the internal transverse directions\footnote{At
least outside the horizon.  It is less clear how to construct an interior
solution by such methods.}.

To see intuitively how such a result can arise, it is useful to 
consider an analogy with flat space electrostatics.  Suppose that 
we have a very deep cylindrical well with conducting walls.  We put some charge
in a bucket at the top and begin to lower the bucket down into the well.
The bucket and winch are arranged so that, instead of lowering the bucket
down the center of the well, the bucket is lowered against one wall 
at some angular coordinate $\theta$ around the cylindrical well. 
In this case, the electric field emerging from the mouth of the well
is not rotationally invariant, but encodes the angular position $\theta$
of the charged bucket.  However, as the bucket is lowered into the well, 
this information will fade from view and the field at the mouth will
become approximately rotationally symmetric.  The well here is 
in analogy with the infinite throat of the incipient extremal
black hole mentioned above. The detailed effects of the singular
fields that arise in the brane case are studied below.

We will see that, in addition to the fivebrane case mentioned above, this
same feature also arises in the construction of BPS black six-branes in ten
dimensional string theory from Kaluza-Klein monopoles, fivebranes, and
onebranes which is familiar from the study of black hole entropy for four
dimensional (four charge) black holes.  We discuss both cases below and
look at potential inhomogeneities in a number of charges.  For each case,
this includes the distribution of onebranes and longitudinal momentum (in
the terminology of \cite{HM1,HM2,HM3}) along the onebranes.  We also study
the distribution of transverse momentum along the onebranes.  The black
fivebrane is addressed in section \ref{fivebrane} and the black sixbrane is
addressed in section \ref{sixbrane}.  We conclude with a few comments in
section \ref{disc}.

\section{Building a three charge black fivebrane}
\label{fivebrane}

	Here we examine a class of BPS saturated black hole solutions of
string theory compactified on $T^5 \times R^5$.  In particular, we will
find it convenient to study ``chiral null models'' so that we may use the
results of \cite{CT}. Our solutions will be generalizations of the models
studied in section 2-4 of \cite{AT2}. Chiral null models, first discussed
in \cite{HT}, are exact solutions of string theory (type IIA, type IIB and
heterotic) in which the only nonzero fields are the metric, the dilaton
$\Phi$, and a Neveu-Schwarz antisymmetric tensor field $H_{ABC}$.  As a
result, our BPS black branes (and related solutions) will be constructed
entirely from Neveu-Schwarz objects.  However, the associated results for
D-branes follow directly by using S-duality.

We now briefly introduce these models since we will use
them explicitly in our study of both the black fivebrane and black sixbrane
solutions.  The 2-dimensional $\sigma$-model Lagrangian coupled to the
above fields is given by
\begin{equation} 
L= (G_{AB} + B_{AB})(X)\pd X^A \bpd X^B + R\Phi(X),
\label{sml}
\end{equation} 
where $G_{AB}(X)$ is the metric, $B_{AB}(X)$ the axion field, $\Phi(X)$ the
dilaton field and $R$ is related to the string worldsheet metric $\gamma$
and its scalar curvature  $\; ^{(2)}R$ by $R = {1 \over 4} ^{(2)}R$.  In the
chiral null model the $\sigma$-model Lagrangian (\ref{sml})
is restricted to take the particular form \cite{CT},
\begin{eqnarray} 
L & = & F(x)\pd u[\bpd v + K(x,u) \bpd u + 2  A_M(x,u) \bpd x^M]\nonumber
 \\ 
 &&+ (G_{MN} +B_{MN})(x)\pd x^M \bpd x^N + R\Phi(x), 
\label{cnm}
\end{eqnarray} 
where $u,v$ are the light-cone coordinates and $x^M$ are coordinates in the
other directions.  We take the topology of the $u,v$ directions to be $S^1
\times \R$, where this $S^1$ is the factor of the $T^5$ mentioned above,
around which the onebranes will be wrapped.  We will refer to the remaining
$T^4$ as the ``transverse internal torus'' below. Note
that these models have $B_{uv} = G_{uv}$ and $B_{uM}=G_{uM}$, and that
${\partial \over {\partial v}}$ is is a null Killing vector. One of the
requirements \cite{CT} for this Lagrangian to be conformally invariant to
all orders in $\alpha'$ is that the functions $F(x), K(x,u), A_i(x,u),
A_a(x,u)$ and $\Phi(x)$ satisfy
\begin{eqnarray} 
-{1\over 2} \nabla^2F^{-1} + \pd^M\psi \pd_MF^{-1} &=&0 \\
-{1\over 2} \nabla^2K  + \pd^M\psi \pd_MK + \pd_u \nabla_MA^M &=&0\\
-{1\over 2} {\hat \nabla}_M \cF^{MN} + \pd_M\psi \cF^{MN} &=&0,
\label{condns}
\end{eqnarray} 
where
\begin{eqnarray} 
{\hat \nabla} \equiv \nabla({\hat \Gamma})\; & ; & \; {\hat \Gamma}^M_{NP}=
\Gamma^M_{NP} + {1 \over 2}H^M_{\; \;\;NP}, \nonumber \\ 
\cF_{MN} = \pd_MA_N -\pd_NA_M \; & ; & \;H_{MNP}=3 \pd_{[M}B_{NP]}
\nonumber \\  
\Phi(x) & = & \psi(x) + {1\over 2} \ln F(x) ,\nonumber
\end{eqnarray}  
with $\nabla_M$ being the connection compatible with the transverse metric
$G_{MN}$. The other requirement is that the lower dimensional $\sigma$-model
$(G_{MN}+B_{MN})\pd x^M \bpd x^N$ be conformal when supplemented by the
dilaton coupling $\psi(x)$.

Such models can be used to describe BPS black fivebranes and related
solutions by breaking the coordinates $x^M$ into two sets $(x^i,y^a)$ where
$x^i (i=1,2,3,4)$ are coordinates in the asymptotically flat directions
transverse to all of the branes and $y^a(a=1,2,3,4)$ are coordinates on the
 transverse internal $T^4$.  For simplicity, we will consider static (i.e.,
$u$-independent) solutions which describe Neveu-Schwarz fivebranes wrapped
around the internal directions (including $u,v$) and Neveu-Schwarz
onebranes wrapped only around the $u,v$ directions.  We allow the branes to
carry longitudinal momentum (momentum in the $u,v$ directions), transverse
internal momentum (in the $y^a$ directions), external linear momentum (in
the $x^i$ directions) and angular momentum associated with the spherical
symmetry in the $x^i$ coordinates.  The various charges need not reside on
the fivebrane and need not be distributed homogeneously around the $T^4$.
Specifically, we consider solutions of the form:

\begin{eqnarray} 
ds^2 & =&  H_1(x,y)^{1/4} H_5(x)^{3/4}[ {{du} \over {H_1(x,y)H_5(x)}} \left(-dv
+ K(x,y)du + 2A_i(x,y)dx^i + 2A_a(x,y)dx^a \right) \nonumber \\
&& + {{dy_a dy^a } \over {H_5(x)}} +  dx_i dx^i ]
\label{outside}
\end{eqnarray} 
\begin{equation}
e^{-2\Phi(x,y)} = {H_1(x,y) \over H_5(x)}
\end{equation}
\begin{equation}
\label{Heqn}
H_{iuv} = H_1^{-2}\pd_i H_1 \qquad H_{auv} = H_1^{-2}\pd_a H_1, \qquad H_{ijk} = -\epsilon_{ijkl}\pd^l H_5
 \end{equation}
in the Einstein frame.  When $A_i=0$  (so that the angular and external
momentum vanishes), these are just the S-dual of  the $u$-independent   
special case of the solutions
constructed in \cite{HM3}.Related solutions were also discussed in
 \cite{AT2}.

The corresponding $\sigma$-model Lagrangian is
\begin{eqnarray} 
L & =&  {1\over H_1(x,y)} \pd u[\bpd v + K(x,y) \bpd u + 2 A_i(x,y) \bpd x^i +
2 A_a(x,y) dy^a] \nonumber \\ 
&& + H_5(x)\pd x_i \bpd x^i + \pd y_a \bpd y^a + B_{ij}(x)\pd x^i \bpd x^j
+ {1 \over 2} R \ln H_5(x), 
\label{Lag}
\end{eqnarray} 
where 
\begin{eqnarray} 
H_{ijk} = 3\pd_{[i}B_{jk]}(x)=-\epsilon_{ijkl}\pd^lH_5(x), \nonumber \\
H_{abc} = H_{abi}=0 = H_{aij}=0. \nonumber
\end{eqnarray} 
Comparing with (\ref{cnm}), $F(x,y)= H_1^{-1}(x,y)$, 
$G_{ij}=H_5(x)\delta_{ij}$, $G_{ia}=0$,
$G_{ab}=\delta_{ab}$.  Substituting into (\ref{condns}), we get the set of equations,
\begin{eqnarray} 
\pd_i^2 H_1(x,y) + H_5(x) \pd_a^2H_1(x,y) & = & 0 \label{h1}\\
\pd_i^2K(x,y)  + H_5(x) \pd_a^2K(x,y)& =& 0 \label{K} \\
\pd_j^2A_i(x,y) + H_5(x)\pd_a^2A_i(x,y) - \pd_i \pd_j A_j(x,y)
-H_5(x)\pd_i\pd_a A_a(x,y) &&\nonumber \\
  - {1 \over 2H_5(x)}  [ \pd_k H_5 (\pd_k
A_i(x,y)-\pd_i A_k(x,y)) - \epsilon_{kjil} \pd_lH_5(x)( \pd_j A_k(x,y) &&
\nonumber \\  
- \pd_k A_j(x,y))]& = & 0 \label{aone}\\ 
\pd_i^2 A_a(x,y) + H_5(x)\pd_b^2A_a(x,y) - \pd_a\pd_iA_i - H_5(x,y) \pd_a
\pd_b A_b & = &0 \label{atwo}. 
\label{ang}
\end{eqnarray}  
Here, the repeated indices are summed even if they are both covariant and
$\pd_i^2=\pd_i\pd_i$. That is, we make implicit use of the metrics
$\delta_{ij}$ and $\delta_{ab}$.

By comparison with, for example, \cite{HM3}, we know that the lower
dimensional $\sigma$-model defined by $G_{MN}$, $B_{MN}$, and $\psi$
will be conformally invariant if we impose

\begin{equation}
\label{five}
\partial_i^2 H_5(x) =0.
\end{equation}
We will assume that the fivebranes are all localized at the origin of the
asymptotically flat coordinates and take $H_5 =  1 + {{r_5^2} \over {r^2}}$
where $r^2 = \sum_{i}x^ix^i$.

Note that the field $H_1$ appears only in equation (\ref{h1}) above.
Thus, having fixed $H_5$, we may
study $H_1$ separately from the remaining fields.  Equation (\ref{h1})
will govern $H_1$ away from any sources, but we would like to include
the sources by considering a properly normalized Green's function for this
equation.  
The normalization may be checked by computing
the total abelian electric charge carried by 
the anti-symmetric tensor field.  The result
shows that the Green's function for a  unit charge satisfies 
\begin{equation} 
\label{green}
\pd_i^2 G({\vec r},{\vec y};{\vec r}_0, {\vec y}_0) + (1 +{r_5^2\over
r^2})\pd_a^2G({\vec r},{\vec y};{\vec r}_0, {\vec y}_0)= -
\delta({\vec r}-{\vec r_0})\delta({\vec y} - {\vec y}_0), 
\end{equation} 
with
\begin{equation} 
\pd_i^2= {1 \over r^3}\pd_r(r^3 \pd_r) + {1\over r^2}{\hat
L}^2(\psi,\theta,\phi), 
\end{equation} 
${\hat L}(\psi,\theta,\phi)$ being the $4$ dimensional angular momentum
operator.  By ``the Green's function for a unit charge,'' we mean that, if
the charge above were smeared out so as to become uniform in the internal
$T^4$ directions, then the solution would be just $H_1 = {1\over 2r^2}$.

The solutions of (\ref{green}) may be studied by expanding $G({\vec r}, {\vec
y};{\vec r}_0, {\vec y}_0)$ in terms of the 3-dimensional spherical
harmonics $D^j_{mn}(\psi,\theta,\phi)$ and the plane wave modes $e^{i{\vec
q}.({\vec y }-{\vec y}_0)}$ in the $T^4$ directions:
\begin{equation} 
G({\vec r}, {\vec y}; {\vec r}_0,  {\vec y}_0) ={1\over V} \sum_{\vec{q},j,m,n}
G_{j \vec q}(r;r_0)D^{*j}_{mn}(\psi_0,\theta_0,\phi_0)D^j_{mn}
(\psi,\theta,\phi) e^{i {\vec q}.({\vec y}- {\vec y}_0)},
\label{exp}
\end{equation}
where ${\hat L}^2(\psi,\theta,\phi)D^j_{mn}(\psi,\theta,\phi)=
-j(j+2)D^j_{mn}(\psi,\theta,\phi)$ and $V$ is the volume of the $T^4$.
This reduces (\ref{green}) to 
\begin{equation} 
{1\over r^3} \pd_r (r^3 \pd_r G_{j \vec q}(r;r_0)) -{1\over r^2}j(j+2) G_{j
\vec q}(r;r_0) - q^2 ( 1 +{r_5^2\over r^2})G_{j \vec q}(r;r_0) = -
{1 \over r^3} \delta(r-r_0).
\end{equation}  
Putting $G_{j \vec q}(r;r_0)={A(r_0)  \over r}Z_{\mu}(r)$, the homogeneous
equation
resembles the modified Bessel equation,  when  $\mu^2=1 +j(j+2) +
q^2r_5^2$. In other words, we find that 
\begin{eqnarray} 
G_{j \vec q}(r;r_0)  =  { q I_{\mu}(qr_0)K_{\mu}(qr)\over r r_0} \; \;
  & {\rm for} & \;  \;r>r_0 \\   
G_{j \vec q}(r;r_0) = {q I_{\mu}(qr)K_{\mu}(qr_0)\over r r_0} \;
  \; & {\rm for} & \; \;  r<r_0,  
\end{eqnarray} 
for $q = |\vec{q}| \neq 0$, and 
\begin{eqnarray} 
G_{j 0} =  {1 \over 2(j+1)}{r_0^j\over r^{j+2}} & \qquad & r>r_0 \\
G_{j 0} =  { 1 \over 2(j+1)}{r^j\over r_0^{j+2}} & \qquad & r<r_0,
\end{eqnarray} 
for $|\vec{q}| = 0$.

Suppose that we examine this solution from the external region $r>r_0$.
Then, as
the onebrane source is brought close to the origin $(r_0 \rightarrow 0)$, we
have, for  $|\vec{q}| \neq 0$,
\begin{equation} 
G_{j \vec q}(r;r_0) \approx   q \left ({q \over 2}\right
)^{\mu} {r_0^{\mu -1} \over \Gamma(\mu +1)} {K_{\mu}(qr)\over r }
\end{equation} 
so that all the $G_{j \vec q}$ vanish except when $|\vec q|=0$ {\it and} $j
=0$.  This means that the only term that survives in the expansion
(\ref{exp}) is the fully homogeneous one.  Note that, so long as the
multipole moments of the source do not increase exponentially, the series
(\ref{exp}) converges geometrically for $r >r_0$ and it is sufficient to
consider each term individually.  Thus, we conclude that the full solution
for $r >0$ becomes homogeneous in the limit.  Although it was clear from
the structure of the equations that the angular dependence would die out as
the onebrane approached $r=0$, it is surprising to find that, even though
there is a non-trivial transverse internal $T^4$ at the origin, the
inhomogeneity in that direction disappears as well.

{}From (\ref{K}) above, it is evident that the equation for the
longitudinal momentum (carried by the $K$-field) is similar, so this charge 
exhibits the same effect. Indeed, if we simplify the discussion by taking both
$A_i$ and $A_a$ to be divergence free\footnote{Note that
such a restriction does not prevent the solution from carrying
angular momentum.} ($\partial_iA_i=0$ and $\partial_a
A_A=0$), then the equation (\ref{atwo})
satisfied by $A_a$ (which carries the transverse momentum) also takes
the form (\ref{green}) and will possess the same feature.  However, because
of the more complicated form of (\ref{aone}), we have not been able to
exhibit a corresponding property of the angular momentum distribution.

\section{Building a four charge black sixbrane}
\label{sixbrane}

Here we consider the compactification of the 10-dimensional theory on $T^6
\times R^4$. Corresponding black hole solutions may be built from
Neveu-Schwarz fivebranes and onebranes together with Kaluza-Klein
monopoles.  We consider 10-dimensional solutions, inspired by the
6-dimensional solutions of \cite{CT,AT2}, 
in which the fivebrane and monopole
charges are located at $r=0$ while the other charges may be distributed
arbitrarily.  Our notation is chosen to match that of section II as closely
as possible, rather than matching that of \cite{CT}.  The fields $H_1$,
$H_5$, and $K$ are related to the distribution of onebranes, fivebranes,
and momentum as above and the field $H_M$ is related to the Kaluza-Klein
monopoles.  In the Einstein frame,

\begin{eqnarray} 
ds^2 &= & H_1^{1 \over 4}(x,y) H_5^{3\over 4}(x) \Bigg[ 
{{du} \over {H_1(x,y) H_5(x)}} du \Bigg( dv +
K(x,y)du + 2
A_w(x,y)a_i(x)dx^i \nonumber \\ && + 2 A_a(x,y)dy^a  
+ 2 A_w(x,y)dw \Bigg)  + {{dw^2} \over {H_M(x)}} + {{dy_ady^a} \over
{H_5(x)}} +
2{{a_i(x)} \over {H_M(x)}} dwdx^i\nonumber \\ 
& & +
\left( H_M^{-1}(x)a_i(x)a_j(x)+ H_M(x)\delta_{ij}
\right) {dx^idx^j} \Bigg],
\end{eqnarray} 
\begin{equation} 
e^{-2\Phi(x,y)}={{H_1(x,y)} \over {H_5(x)}},
\qquad H_{wij}=\epsilon_{ijl}\pd^l H_5,
\end{equation}
where the monopole is in the $w$ direction and the labels $a=1,2,3,4$ span
the transverse internal $T^4$ directions while $i=1,2,3$, span the spatial
coordinates.  The associated $\sigma$-model Lagrangian is
\begin{eqnarray} 
L & = & H_1^{-1}(x,y)\pd u( \bpd v + K(x,y) \bpd u + 2 A_w(x,y)[\bpd w +
a_i(x) \bpd x^i] +2 A_a(x,y)\bpd y^a) \nonumber \\  
& & + {1 \over 2} R \ln H_1^{-1}(x,y ) + L_{\perp},
\end{eqnarray}
with  
\begin{eqnarray} 
L_{\perp} & =&  \pd y_a \bpd y^a + H_5(x)H_M^{-1}(x)
[\pd w + a_i(x) \pd x^i][\bpd w
+ a_j(x) \bpd x^j]  \nonumber \\ 
& & + H_5(x) H_M(x) \pd x^i \bpd x^i + b_i(x)[\pd w  \bpd x^i - \bpd w
\pd x^i] + R\psi(x). 
\label{sol}
\end{eqnarray} 
Note that there is an additional translational 
Killing vector ${\pd \over \pd w}$ in the
monopole direction as well. Moreover, we follow \cite{CT} in taking the
functions $H_5(x), H_M(x), b_i(x)$ and $a_i(x)$ to satisfy
\begin{eqnarray} 
\pd_i^2 H_5 (x)=0 & ; &   \pd_i^2 H_M(x) = 0 \label{fk} \\
\pd_ib_j(x) -\pd_jb_i(x)= -\epsilon_{ijl}\pd^l H_5(x) & ; & \pd_ia_j(x) -\pd_ja_i(x)=
-\epsilon_{ijl}\pd^l H_M(x)\\
\psi(x) &=& {1\over 2}\ln H_5(x)
\end{eqnarray} 
so that $L_\perp$ is conformally invariant.  The
conditions   (\ref{condns}) of the chiral null model
reduce to\footnote{This computation was done using MathTensor.} 
\begin{eqnarray} 
\pd_i^2 H_1 + H_5H_M\pd_{a}^2H_1& = &0 \label{F} \\
\pd_i^2 K + H_5 H_M\pd_{a}^2K &= & 0\label{K4}\\
A_w[H_M^{-2}a_ia_j\pd_i H_5 \pd_j H_M^{-1} + \pd_i H_5 \pd_i H_M^{-1} +
H_5 H_M^{-1} \epsilon_{ijl}\pd_ia_l\pd_j H_M^{-1} & &
\nonumber\\ 
+ H_M^{-1} \epsilon_{ijl}a_i\pd_j H_5 \pd_l H_M^{-1}] 
+ H_M^{-2} \epsilon_{ijl}a_i\pd_j H_5 \pd_l A_w + H_M^{-1} \pd_i H_5 
\pd_i A_w && \nonumber \\
+ H_5 \pd_i H_M^{-1} \pd_iA_w + H_5 H_M^{-1} \epsilon_{ijl}a_i\pd_j H_M^{-1} \pd_lA_w - H_5 H_M^{-1}
 \pd_i^2 A_w && \nonumber \\ - H_5^2H_M^{-2} a_i\pd_i\pd_aA_a 
- H_5^2 \pd_a^2 A_w &=&0 \label{Aonea}\\
-H_M^{-1}\pd_i^2A_a - H_5 \pd_{b}^2A_a + H_5
\pd_a \pd_{b}A_{b} &=&0\label{Aoneb}\\
A_w[\epsilon_{ijl}\pd_j H_M^{-1} \pd_l H_5 - H_M^{-1} a_j\pd_j H_5 \pd_i
H_M^{-1}] - H_M^{-1} \epsilon_{ijl}
\pd_j H_5 \pd_l A_w && \nonumber \\
- H_5 \epsilon_{ijl}\pd_j H_M^{-1} \pd_l A_w + H_5^2 H_M^{-1} \pd_i
\pd_a A_a &=&0 \label{Aonec}.
\end{eqnarray} 
Here the index $b$ refers to the $T^4$ internal directions and is summed
over.  Since the fivebrane and monopole charges are to reside at $r=0$,
using (\ref{fk}), we take $H_5 = 1 + r_5/r$ and $H_M = 1 + r_{M}/r$.  As
before, we will focus on the field associated with the onebrane
distribution ($H_1$).  This time, however, we will only be able to discuss
the limiting case in which the onebrane charge is located near $r=0$.
Then, to leading order the Green's function for equation (\ref{F})
satisfies
\begin{equation} 
\label{g2}
(\pd_i^2  + {r_5r_{M}\over r^2}\pd_{a}^2)G({\vec r},{\vec y}; {\vec
r}_0,{\vec y}_0) \approx
-\delta({\vec r}-{\vec r}_0)\delta({\vec y}-{\vec y}_0).
\label{FF}
\end{equation} 
Again, the norm can be checked by computing the total electric charge for
such a solution. Expanding $G({\vec r},{\vec y}; {\vec r}_0,{\vec y}_0)$ as in (\ref{exp}),
we get  
\begin{equation} 
(\pd_r(r^2\pd_r)-l(l+1) -q^2r_5r_{M}) G_{\vec q l}(r;r_0) \approx -\delta({r}-{r}_0),
\end{equation} 
which has  solutions,
\begin{eqnarray} 
G_{\vec q l} =  {1\over \sqrt{1+s^2 }{\sqrt{r r_0}} }\left ({r\over
r_0}\right )^{{\sqrt{1+s^2} \over 2}}  \;  \; &{\rm for}& \; \;   r<r_0
\nonumber \\
G_{\vec q l} = {1\over \sqrt{1+s^2 }{\sqrt{r r_0}} }\left({r_0\over
r}\right )^{{\sqrt{1+s^2} \over 2}} \;  \;  & {\rm for}& \;  \;  r>r_0  ,
\end{eqnarray} 
where $s^2=4(l(l+1) + q^2)$. Examining the solution from the exterior
region, we see that as $r_0 \rightarrow 0$, the $G_{\vec q l}(r;r_0)$
vanish except when $s^2=0$, which is only satisfied if $l=0$ {\it and}
$|\vec q| =0$.  A similar analysis follows from (\ref{K4}) for the
longitudinal momentum carried by the field $K$. Here then, is another case
in which the onebrane charge and the longitudinal momentum distributions
become homogeneous when the charge is moved to $r=0$.

A similar examination of the monopole charge and the
$w$-momentum is, on the other hand, not so straightforward as we can
see from (\ref{Aonea}), (\ref{Aoneb}) and (\ref{Aonec}). However, if we set
the $w$-momentum to zero ($A_w=0$) and impose a divergence free
condition $\partial_a A_a=0$,  we find that 
(\ref{Aonea}), (\ref{Aoneb}), and 
(\ref{Aonec}) reduce to a single equation of the form studied above:
\begin{equation} 
\pd_i^2 A_a + H_5 H_M\pd_{b}^2 A_a  
=0. \label{third} 
\end{equation} 
Thus,  the transverse internal momentum
also becomes translationally invariant when a fixed distribution of 
sources is moved to $r=0$.

\section{Discussion}
\label{disc}

We have seen that, when collections of branes are brought close together,
the information about the distribution of charges in the internal
transverse directions will be obscured from an observer who remains a fixed
distance away.  So long as the intrinsic moments of the charge
distributions are held fixed, the multipole moments that are measured from
infinity will vanish in the limit as the branes are brought together.  We
have studied this in detail for the case of bringing onebranes with
longitudinal and transverse internal momentum together with a fivebrane in
string theory (type IIA, IIB and heterotic), and for the case of bringing
such onebranes together with an already bound state of Kaluza-Klein
monopoles and fivebranes in string theory. Although they are of a slightly
different form, we also note that the solutions of \cite{ITY} with strings
localized on fivebranes cannot be formed from a finite collection of
branes.
This can be seen
from the fact that for such solutions, 
even after compactifying the translationally invariant
directions, the integral of $H_1(x,y)-1$ around a sphere at infinity (i.e., 
the total onebrane charge) diverges.
The solutions studied in the
present paper were static, but from the results of \cite{HM3} we see that,
at least for the fivebrane case when $A_i=0$, a trivial extension allows an
arbitrary $u$ dependence in the longitudinal and transverse waves so long
as the transverse wave is divergence free.  Having seen this effect in
these contexts, it is natural to assume that it occurs more generally.  In
principle, an analysis of equations (\ref{aone}) and (\ref{atwo}) would
determine if angular momentum on a BPS black fivebrane and other waves
associated with the asymptotically flat directions also behave in this way.

A moment should be taken to comment on our bringing the charges together
``while holding the intrinsic charge distribution fixed."  Clearly, the
distribution need only be fixed in the limit as the charges are brought
together.  However, the reader may wonder just which limiting distributions
are allowed and which are not.  Recall that our analysis in sections II and
III worked from the multipole moments of the charge distribution around the
transverse internal $T^4$.  All that we require is that these moments
increase less than exponentially with the mode number $q$ so that the
convergence of the series (\ref{exp}) can be controlled.  As a result, a
$\delta$-function (which would describe, for example, a single localized
onebrane) and in fact all tempered distributions are included in our
treatment.  Thus, the only way that an inhomogeneous solution might be
obtained when the charges are brought together is if they approach an
arrangement more singular than any tempered distribution.

Although our result may seem surprising at first, in retrospect it should
perhaps have been expected.  This can be seen from an analogy with
Einstein-Maxwell
theory.  Consider what happens when we bring a fixed distribution
of electric charge near a Schwarzschild black hole.
Let us suppose that the charge distribution is localized on a scale
much smaller than the black hole horizon.  If we allow the electric
charge to fall into the black hole then, due to the ``no hair'' theorems
\cite{schw} the electric field at any value of the radial coordinate
outside the horizon rapidly becomes spherically symmetric even though
the charge does not cross the horizon until $t= \infty$ in Schwarzschild
coordinates.   The case of the BPS branes is, however, 
slightly different due to the existence of static solutions in which
the onebranes remain at a finite separation from the fivebrane.
In the Einstein-Maxwell example above, in order to create a static
solution one would need to add some extra stress-energy representing, for
example, a string to hold the charge up and keep it from falling into the
black hole.  In
a sequence of static solutions in which the charge 
is lowered to the horizon, this stress energy would diverge with the
increasing proper acceleration of the charge.  Thus, the limit 
of a sequence of static solutions would in fact have a singular
horizon. 

A better analogy would be to consider the case
of a small extremal black hole approaching a large extremal black hole.
In this case, no extra forces are needed and the electric field rapidly
becomes spherically symmetric as the horizons approach each other. 
This can be seen directly from the Majumdar-Papapetrou solutions \cite{Maj,Pap}.
The same spherical symmetry is also broken when our branes (from
sections II and III) are separated, and
it is restored in the same way when they come together.
However, because isolated branes are point
objects in the asymptotically flat directions, it comes as no surprise
that spherical symmetry is restored when the two objects merge.

Exactly the same argument can be applied to the singularities discussed
in \cite{KMR} associated with the nonspherical longitudinal waves.
Suppose that we attempt to assemble a nonspherical wave by bringing together
in a spherically asymmetric manner
a number of objects (either localized branes or black branes), each of which
individually is a spherically symmetric\footnote{One can make this
precise by considering black branes and requiring the horizon of each to
be symmetric under the appropriate $SO(n)$ group.}.  Then, despite our
efforts, the wave carried by the final merged object will be spherically
symmetric.  Mathematically, this case is exactly equivalent 
to the merger of extremal black holes just discussed in the context
of the Majumdar-Papapetrou solutions.

In addition to the black hole cases already considered above, it is
plausible that a similar feature arises for the various intersecting brane
solutions (such as those in \cite{BC,BMM,MM,BLL}).  While such cases do not
form an infinitely deep throat, the singular fields near the branes may
well produce a similar effect.  All of these cases seem worthy of
investigation.

\acknowledgements We would like to thank Gary Horowitz for discussions of
the chiral null models and the various solutions.  We also thank
A.A. Tseytlin for comments on an earlier draft. SS would like to thank the
Physics Department at Syracuse University for its hospitality. This work
was supported in part by the Inter University Center for Astronomy and
Astrophysics, Pune, India, NSF grant PHY-9722362, and funds provided by
Syracuse University.

\end{document}